\documentclass{appolb}
\usepackage{graphicx}
\usepackage[square,sort&compress,numbers]{natbib}
\begin{document}
% \eqsec  % uncomment this line to get equations numbered by (sec.num)
\title{HERA Inclusive Neutral and Charged Current Cross Sections and a New PDF Fit, HERAPDF\,2.0%
\thanks{Presented at EDS Blois 2015 Conference for the H1 and ZEUS Collaborations}%
% you can use '\\' to break lines
}
\author{Zhiqing Zhang
\address{Laboratoire de l'Acc\'el\'erateur Lin\'eaire, Univ.\ Paris-Sud et IN2P3/CNRS, France}
%\\
%{Third Author of different affiliation
%}
%the Name(s) of other Author(s)
%\address{affiliation}
}
\maketitle
\begin{abstract}
In this talk, I present the brand new results from the H1 and ZEUS Collaborations on the combination of all previously published inclusive deep inelastic cross sections at HERA for neutral and charged current $e^\pm p$ scattering for zero beam polarisation and the corresponding parton distributions functions, HERAPDF\,2.0, at up to next-to-next-to-leading order (NNLO). The results also include a new precise determination at next-to-leading order (NLO) of the strong coupling constant $\alpha_s(M^2_Z)=0.1184\pm 0.0016$ (excluding scale uncertainties) based on a simultaneous fit to the combined inclusive cross section data and jet production data. 
\end{abstract}
  
\section{Introduction}
The $e^\pm p$ collider, HERA, used to be the largest electron microscope of the world, where the inclusive deep inelastic scattering (DIS) scattering of neutral and charged current (NC and CC) interactions has been studied over an unprecedented large kinematic region of Bjorken $x$ and negative four-momentum-transfer squared $Q^2$.

Based on the combined inclusive NC and CC cross sections measured by the H1 and ZEUS experiments using data taken from 1992 to 2000 in phase\,1 (HERA\,I), a set of parton distribution functions (PDFs), HERAPDF\,1.0, was previously obtained~\cite{herapdf1}. These data have also been the primary input for all modern PDF sets such as CT10~\cite{ct10}, MMHT2014~\cite{mmht} and NNPDFs~\cite{nnpdf}.

In phase\,2 (HERA\,II), from 2002 to 2007, the integrated luminosity of the $e^+p$ ($e^-p$) collision amounts to about 150\,pb$^{-1}$ (235\,pb$^{-1}$), representing an increase by a factor of 1.5 (15) over that of the corresponding number at HERA\,I. In addition, the $e^\pm$ beams were longitudinally polarised at HERA\,II. However for the derivation of HERAPDF\,2.0, the HERA\,II data were corrected to zero beam polarisation so that a full combination with the unpolarised cross sections from HERA\,I could be made.

This writeup is organised as follows. In Sec.~\ref{sec:combination}, the data combination is briefly described. In Sec.~\ref{sec:pdf}, the QCD analysis to extract PDFs from the combined inclusive data is presented, followed by a summary in Sec.~\ref{sec:summary}. Full details of the analysis are given in~\cite{herapdf2}.

\section{Data combination}\label{sec:combination}
In total, 41 data sets have been used in the combination, of which 21 are from HERA\,I, taken mainly at nominal proton beam energy ($E_p$) of 920\,GeV but also at 820\,GeV, and 12 from HERA\,II are at 920\,GeV and 4 each at 575 and 460\,GeV, corresponding to centre-of-mass energies ($\sqrt{s}$) of 320, 300, 251 and 225\,GeV.

A total number of 2927 individual data points is transformed and combined into a set of 1307 common grid points. The transformation is performed based on predictions at high $Q^2$ using an in-situ NLO QCD fit to the data and at low $Q^2$ a fractal model~\cite{fmodel} based fit. The combination is made by averaging the transformed data points based on a $\chi^2$ minimisation method~\cite{average}, assuming that there is only one correct value for the cross section of each process at each point of the phase space. The correlation within a data set and between different data sets is taken into account in the combination.

\begin{figure}[htb]
%\begin{center}
\includegraphics[width=0.495\textwidth]{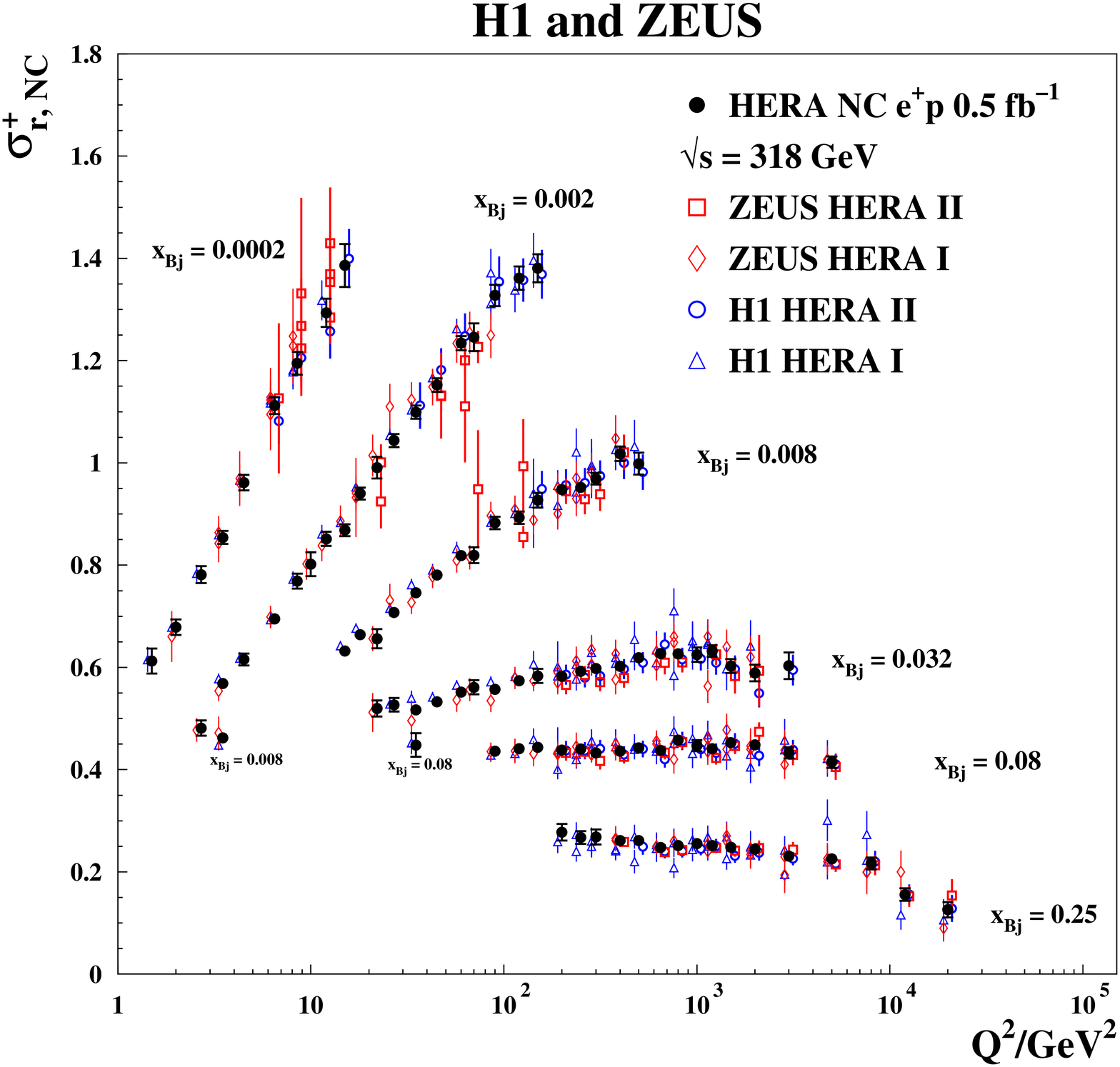}
\includegraphics[width=0.545\textwidth]{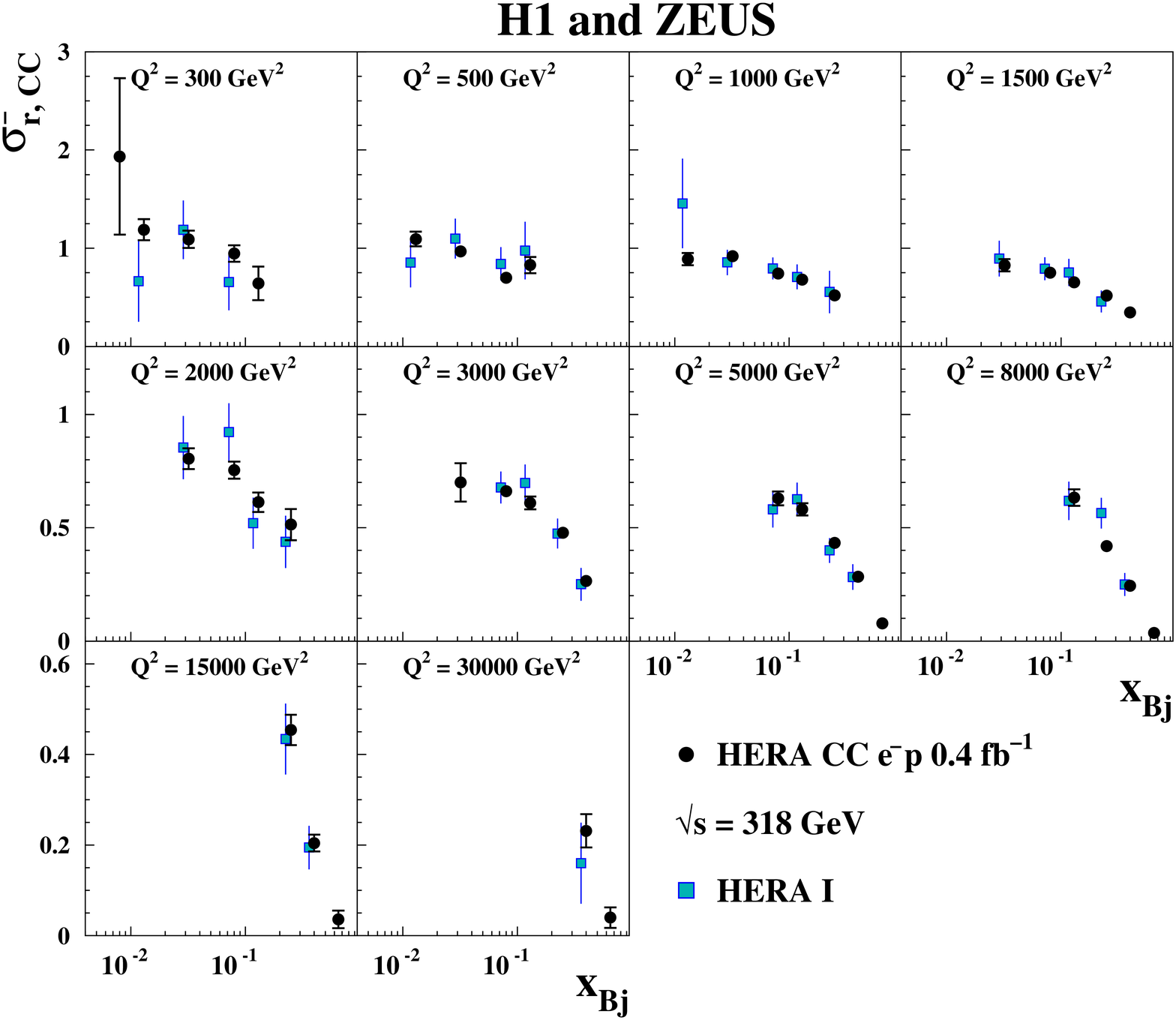}
%\end{center}
\caption{Left: some of the selected H1 and ZEUS individual NC $e^+p$ cross section measurements in HERA\,I and II in comparison with the corresponding combined data. Right: the combined CC $e^-p$ cross sections from HERA\,I in comparison with the full combination.}
\label{fig:datacomb}
\end{figure}

The combination has significantly improved not only the statistical precision but also the systematic one since H1 and ZEUS used different reconstruction methods. Also for certain regions of the phase space, where one of the two experiments has superior precision compared to the other, the less precise measurement is fitted to the more precise measurement, with a simultaneous reduction of the correlated systematic uncertainty. This reduction propagates to the other points, including those which are based solely on the measurement from the less precise experiment. Two examples are shown in Fig.~\ref{fig:datacomb}. 
The first example presents the reduction of the uncertainty from some of the selected individual NC $e^+p$ cross sections to the combined one and the second one the reduction from the combined CC $e^-p$ HERA\,I data to the full combination. The first example also shows the strong scaling violation effect in the NC data which gives rise to the sensitivity to constrain the gluon distribution function of the proton. The data at different $E_p$ and thus $\sqrt{s}$ allow a direct measurement of the longitudinal structure function $F_L$ providing further sensitivity to the gluon distributions.

\section{HERAPDF\,2.0}\label{sec:pdf}
The availability of precision NC and CC cross sections over a large phase space allows HERAPDF to be based on $ep$ scattering data only and makes HERAPDF independent of any nuclear corrections.
The new HERAPDF\,2.0 is obtained following the same framework used for HERAPDF\,1.0~\cite{herapdf1}. Five PDFs $xu_v(x)$, $xd_v(x)$, $x\overline{U}(x)$, $x\overline{D}(x)$ and $xg(x)$ are parameterised at an initial scale $\mu^2_{f_0}$, taken to be 1.9\,GeV$^2$, in a generic form $xf(s)=Ax^B(1-x)^C(1+Dx+Ex^2)$ allowing $xg$ to have an extra term $-A^\prime_gx^{B^\prime_g}(1-x)^{C^\prime_g}$ for more flexibility~\cite{herapdf2}. The theoretical predictions of the NC and CC cross sections are obtained by convoluting the PDFs with coefficient functions at LO, NLO or NNLO in the $\overline{\rm MS}$ scheme. The evolution in $Q^2$ of the PDFs is obtained from the DGLAP evolution equations~\cite{dglap}. The renormalisation and factorisation scales are chosen to be $\mu^2_r=\mu^2_f=Q^2$. The heavy quarks are treated in the general-mass variable-flavour-number scheme, RTOPT~\cite{rtopt}, for the NC predictions. For the CC predictions, the zero-mass approximation is used, since all HERA CC data are well beyond the $b$-quark mass scale.

\begin{figure}[htb]
%\begin{center}
\includegraphics[width=0.505\textwidth]{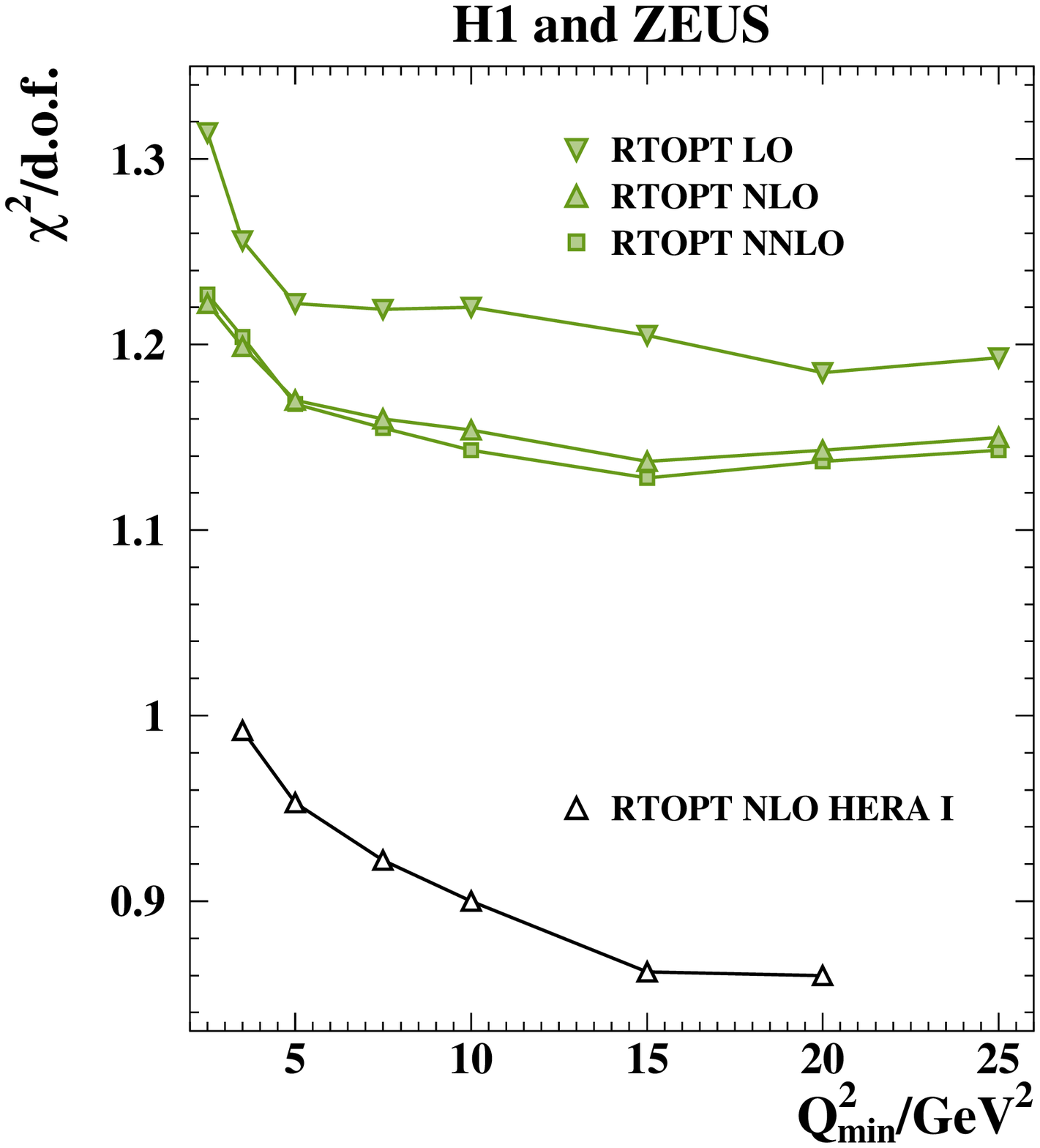}
\includegraphics[width=0.495\textwidth]{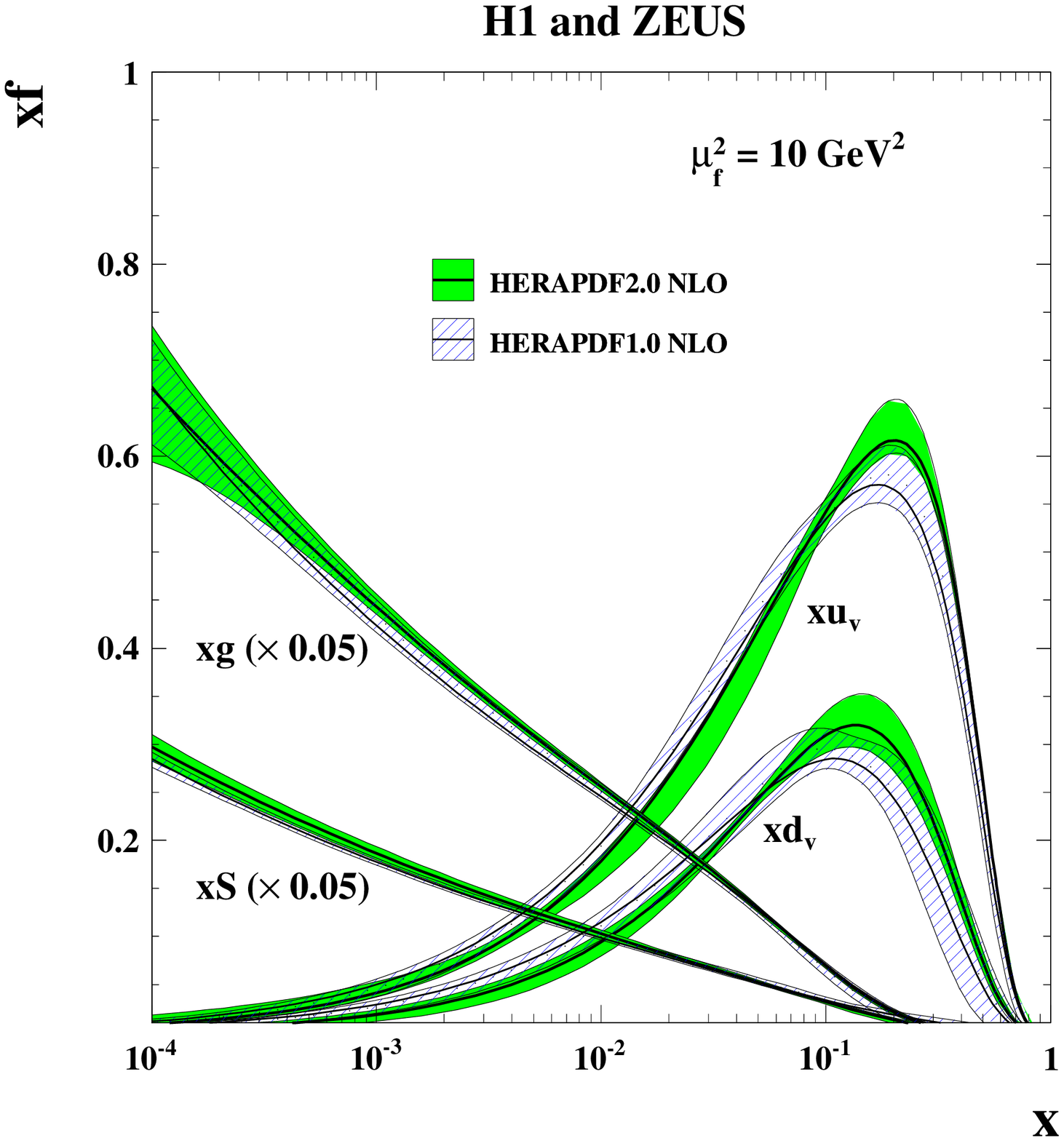}
\includegraphics[width=0.54\textwidth]{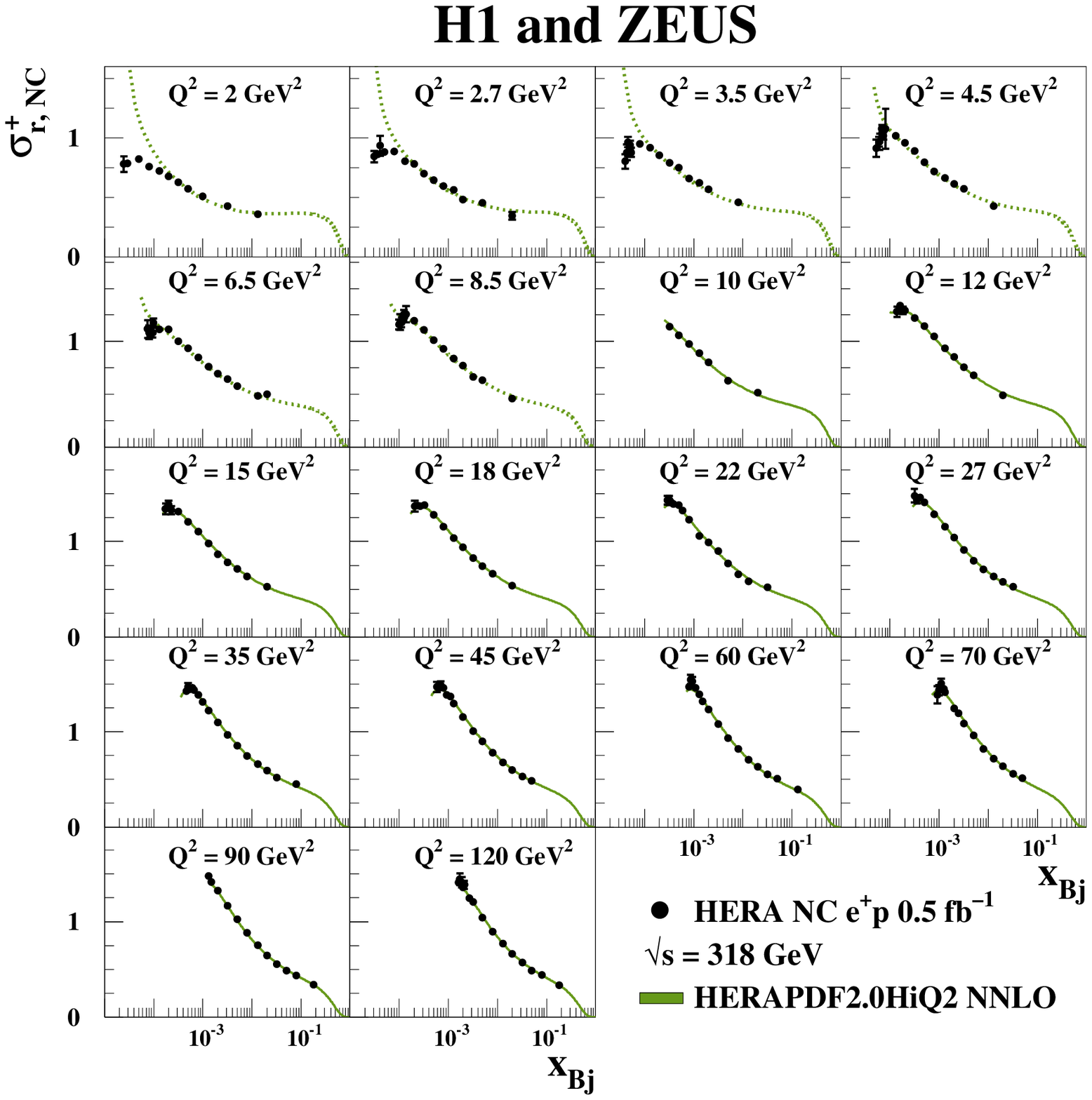}
\includegraphics[width=0.515\textwidth]{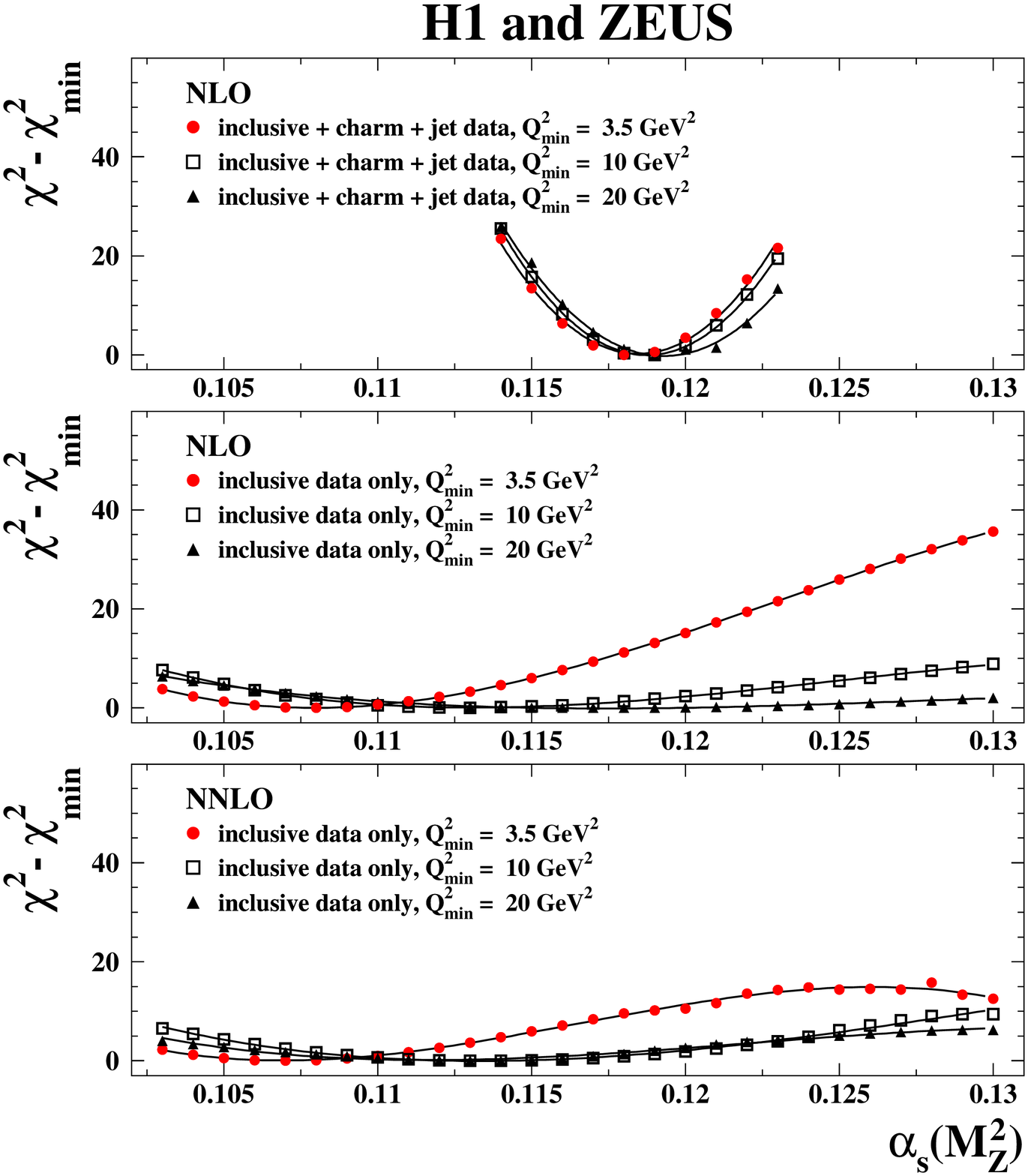}
%\end{center}
\caption{Top left: the dependence of $\chi^2$ per degree of freedom (d.o.f.) on $Q^2_{\rm min}$ of LO, NLO and NNLO fits to the combined data. Top right: comparison of NLO PDFs $xu_v$, $xd_v$, $xS=2x(\overline{U}+\overline{D})$ and $xg$ between HERAPDF\,2.0 and 1.0.
Bottom left: The combined low-$Q^2$ NC $e^+p$ cross section data in comparison with predictions from high $Q^2$ version of NNLO HERAPDF\,2.0. Dotted lines indicate extrapolation into kinematic regions not included in the fit. Bottom right: $\chi^2-\chi^2_{\rm min}$ vs. $\alpha_s(M^2_Z)$ for fits with different $Q^2_{\rm min}$ using (upper part) inclusive charm and jet production at NLO, (middle part) inclusive $ep$ scattering data only at NLO and (lower part) inclusive $ep$ scattering data only at NNLO.}
\label{fig:fit}
\end{figure}

Only cross section data above a minimum scale $Q^2_{\rm min}=3.5\,{\rm GeV}^2$ are used in the fit. The fit is performed by minimising a $\chi^2$ function similar to HERAPDF\,1.0 but with an additional logarithmic term introduced in~\cite{h1pdf2}. The $\chi^2$ value per degree of freedom (d.o.f.) as a function of $Q^2_{\rm min}$ is shown in Fig.~\ref{fig:fit} (top left). The large $\chi^2$/d.o.f. is investigated revealing tension between the QCD predictions and the data at both low and high $Q^2$. The  HERAPDF\,2.0 PDFs corresponding to choosing $Q^2_{\rm min}=3.5\,{\rm GeV}^2$  are shown for $\mu^2_f=10\,{\rm GeV}^2$ in Fig.~\ref{fig:fit} (top right) in comparison with that from HERAPDF\,1.0. The new PDFs have substantially better precision in particular at high $x$ and the new $u$ and $d$ valence quarks are slightly harder but both sets are consistent within the uncertainty bands which include the experimental, model and parameterisation uncertainties~\cite{herapdf2}.

Given the improved $\chi^2$/d.o.f. at higher $Q^2_{\rm min}$, a HERAPDF\,2.0 variant with $Q^2_{\min}=10\,{\rm GeV}^2$, HERAPDF\,2.0HIQ2, is obtained, whose predictions at NNLO are compared with the combined NC $e^+p$ data for $Q^2$ below 120\,GeV$^2$ in Fig.~\ref{fig:fit} (bottom left). Good agreement is observed in the fitted $Q^2$ region. The extrapolation to the low $Q^2$ and $x$ shows, however, clear deviation between the prediction and the low $x/Q^2$ data, which may indicate that one needs to add further contributions (e.g.\ resummation of $\ln(1/x)$ terms) beyond the fixed NNLO order for the low $x/Q^2$ predictions.

The gluon distribution is known to be strongly correlated with the strong coupling constant $\alpha_s$. Using the inclusive NC and CC cross section data is not able to simultaneously determine both. This is shown in Fig.~\ref{fig:fit} (bottom right). This is why the $\alpha_s(M^2_Z)$ value is fixed for HERAPDF\,2.0 to 0.118 at both NLO and NNLO and to 0.130 at LO. The same figure also shows that by adding jet data (combined charm data is also included, however its main effect is to constrain the pole mass of the charm quark to be  1.47 (1.43)\,GeV in the NLO (NNLO) fit), a simultaneous fit of $\alpha_s(M^2_Z)$ and PDFs becomes possible. The resulting $\alpha_s(M^2_Z)$ at NLO amounts to $0.1184\pm 0.0009({\rm exp}) \pm 0.0005({\rm model/parameterisation}) \pm 0.0012({\rm hadronisation}) ^{+0.0037}_{-0.0030}({\rm scale})$, which is competitive in precision with other NLO determinations. 
Note that the jet data cannot be included in an NNLO fit as its prediction in DIS has not been calculated to NNLO.

\section{Summary}\label{sec:summary}
The full combination of the inclusive NC and CC cross sections published previously by H1 and ZEUS at HERA\,I and II is finally ready. The combined data, covering six orders of magnitude in both $x$ and $Q^2$ and reaching an unprecedented precision with typical values of below 0.5\% for medium $Q^2$ region at the nominal centre-of-mass energy, represent a major legacy of HERA. %The data provide a timely and important input for improving PDF uncertainties needed for precise and reliable prediction for the LHC Run\,2.

A new PDF set, HERAPDF\,2.0 at LO, NLO and NNLO, is derived with a QCD fit to the combined inclusive data for $Q^2$ above 3.5\,GeV$^2$. The new PDFs have smaller experimental uncertainties over that of HERAPDF\,1.0.  

By including additional jet and charmed data from HERA, a precision determination of $\alpha_s(M^2_Z)=0.1184\pm 0.0016$ at NLO (excluding the dominant scale uncertainties) in obtained. The value is in excellent agreement with the world average value of 0.1185~\cite{pdg}.


\begin{thebibliography}{99}

\bibitem{herapdf1}
F.~D.~Aaron {\it et al.} [H1 and ZEUS Collaborations],
  %``Combined Measurement and QCD Analysis of the Inclusive e+- p Scattering Cross Sections at HERA,''
  JHEP {\bf 1001} (2010) 109
  [arXiv:0911.0884 [hep-ex]].
  %%CITATION = ARXIV:0911.0884;%%
  %631 citations counted in INSPIRE as of 17 sept. 2015

\bibitem{ct10}
H.~L.~Lai, M.~Guzzi, J.~Huston, Z.~Li, P.~M.~Nadolsky, J.~Pumplin and C.-P.~Yuan,
  %``New parton distributions for collider physics,''
  Phys.\ Rev.\ D {\bf 82} (2010) 074024
  [arXiv:1007.2241 [hep-ph]].
  %%CITATION = ARXIV:1007.2241;%%
  %1431 citations counted in INSPIRE as of 17 sept. 2015
  
\bibitem{mmht}
L.~A.~Harland-Lang, A.~D.~Martin, P.~Motylinski and R.~S.~Thorne,
  %``Parton distributions in the LHC era: MMHT 2014 PDFs,''
  Eur.\ Phys.\ J.\ C {\bf 75} (2015) 5,  204
  [arXiv:1412.3989 [hep-ph]].
  %%CITATION = ARXIV:1412.3989;%%
  %48 citations counted in INSPIRE as of 18 sept. 2015
  
\bibitem{nnpdf}
R.~D.~Ball, L.~Del Debbio, S.~Forte, A.~Guffanti, J.~I.~Latorre, J.~Rojo and M.~Ubiali,
  %``A first unbiased global NLO determination of parton distributions and their uncertainties,''
  Nucl.\ Phys.\ B {\bf 838} (2010) 136
  [arXiv:1002.4407 [hep-ph]].
  %%CITATION = ARXIV:1002.4407;%%
  %377 citations counted in INSPIRE as of 17 sept. 2015

\bibitem{herapdf2}
H.~Abramowicz {\it et al.} [H1 and ZEUS Collaborations],
  %``Combination of Measurements of Inclusive Deep Inelastic $e^{\pm}p$ Scattering Cross Sections and QCD Analysis of HERA Data,''
  arXiv:1506.06042 [hep-ex].
  %%CITATION = ARXIV:1506.06042;%%
  %13 citations counted in INSPIRE as of 18 sept. 2015
  
\bibitem{fmodel}
T.~Lastovicka,
  %``Selfsimilar properties of the proton structure at low x,''
  Eur.\ Phys.\ J.\ C {\bf 24} (2002) 529
  [hep-ph/0203260].
  %%CITATION = HEP-PH/0203260;%%
  %42 citations counted in INSPIRE as of 18 sept. 2015

\bibitem{average}
F.~D.~Aaron {\it et al.} [H1 Collaboration],
  %``Measurement of the Inclusive ep Scattering Cross Section at Low Q^2 and x at HERA,''
  Eur.\ Phys.\ J.\ C {\bf 63} (2009) 625
  [arXiv:0904.0929 [hep-ex]].
  %%CITATION = ARXIV:0904.0929;%%
  %101 citations counted in INSPIRE as of 18 sept. 2015

\bibitem{dglap}
V.~N.~Gribov and L.~N.~Lipatov, Sov.\ J.\ Nucl.\ Phys.\ {\bf 15} (1972) 438, 675; L.~N.~Lipatov, Sov.\ J.\ Nucl.\ Phys.\ {\bf 20} (1975) 94; Y.~L.~Dokshitzer, Sov.\ Phys.\ JETP {\bf 46} (1977) 641; G.~Altarelli and G.~Parisi, Nucl.\ Phys.\ B {\bf 126} (1977) 298.

\bibitem{rtopt}
R.~S.~Thorne and R.~G.~Roberts, Phys.\ Rev.\ D {\bf 57} (1998)  6871 [hep-ph/9709442]; R.~S.~Thorne, Phys.\ Rev.\ D {\bf 73} (2006) 054019 [hep-ph/0601245]; {\bf 86} (2012) 074017 [arXiv:1201.6180 [hep-ph]].

\bibitem{h1pdf2}
F.~Aaron {\it et al.} [H1 Collaboration], JHEP {\bf 09} (2012) 061 [arXiv:1206.7007 [hep-ex]].

\bibitem{ct10nnlo}
J.~Gao {\it et al.},
  %``CT10 next-to-next-to-leading order global analysis of QCD,''
  Phys.\ Rev.\ D {\bf 89} (2014) 033009
  [arXiv:1302.6246 [hep-ph]].
  %%CITATION = ARXIV:1302.6246;%%
  %242 citations counted in INSPIRE as of 18 sept. 2015
  
\bibitem{nnpdf3.0}
R.~D.~Ball {\it et al.} [NNPDF Collaboration],
  %``Parton distributions for the LHC Run II,''
  JHEP {\bf 1504} (2015) 040
  [arXiv:1410.8849 [hep-ph]].
  %%CITATION = ARXIV:1410.8849;%%
  %82 citations counted in INSPIRE as of 18 sept. 2015

\bibitem{pdg}
K.~A.~Olive et al. (Particle Data Group), Chinese Physics C {\bf 38} (2014) 090001.
  
\end{thebibliography}
\end{document}